\documentclass [12pt] {article}
\usepackage{a4}

  \def\pp{{\mathchoice
          {
              \kern 1pt%
              \raise 1pt
              \vbox{\hrule width5pt height0.4pt depth0pt
                    \kern -2pt
                    \hbox{\kern 2.3pt
                          \vrule width0.4pt height6pt depth0pt
                          }
                    \kern -2pt
                    \hrule width5pt height0.4pt depth0pt}%
                    \kern 1pt
           }
            {
              \kern 1pt%
              \raise 1pt
              \vbox{\hrule width4.3pt height0.4pt depth0pt
                    \kern -1.8pt
                    \hbox{\kern 1.95pt
                          \vrule width0.4pt height5.4pt depth0pt
                          }
                    \kern -1.8pt
                    \hrule width4.3pt height0.4pt depth0pt}%
                    \kern 1pt
            }
            {
              \kern 0.5pt%
              \raise 1pt
              \vbox{\hrule width4.0pt height0.3pt depth0pt
                    \kern -1.9pt  
                    \hbox{\kern 1.85pt
                          \vrule width0.3pt height5.7pt depth0pt
                          }
                    \kern -1.9pt
                    \hrule width4.0pt height0.3pt depth0pt}%
                    \kern 0.5pt
            }
            {
              \kern 0.5pt%
              \raise 1pt
              \vbox{\hrule width3.6pt height0.3pt depth0pt
                    \kern -1.5pt
                    \hbox{\kern 1.65pt
                          \vrule width0.3pt height4.5pt depth0pt
                          }
                    \kern -1.5pt
                    \hrule width3.6pt height0.3pt depth0pt}%
                    \kern 0.5pt
            }
        }}

  \def\mm{{\mathchoice
                       {
                             \kern 1pt
               \raise 1pt    \vbox{\hrule width5pt height0.4pt depth0pt
                                  \kern 2pt
                                  \hrule width5pt height0.4pt depth0pt}
                             \kern 1pt}
                       {
                            \kern 1pt
               \raise 1pt \vbox{\hrule width4.3pt height0.4pt depth0pt
                                  \kern 1.8pt
                                  \hrule width4.3pt height0.4pt depth0pt}
                             \kern 1pt}
                       {
                            \kern 0.5pt
               \raise 1pt
                            \vbox{\hrule width4.0pt height0.3pt depth0pt
                                  \kern 1.9pt
                                  \hrule width4.0pt height0.3pt depth0pt}
                            \kern 1pt}
                       {
                           \kern 0.5pt
             \raise 1pt  \vbox{\hrule width3.6pt height0.3pt depth0pt
                                  \kern 1.5pt
                                  \hrule width3.6pt height0.3pt depth0pt}
                           \kern 0.5pt}
                       }}

\catcode`@=11
\def\un#1{\relax\ifmmode\@@underline#1\else
        $\@@underline{\hbox{#1}}$\relax\fi}
\catcode`@=12

\let\du=\du                     

\def\a{\alpha}
\def\b{\beta}

\def\f{\phi}
\def\g{\gamma}
\def\h{\eta}

\def\j{\psi}

\def\m{\mu}

\def\p{\pi}
\def\q{\theta}
\def\r{\rho}
\def\s{\sigma}
\def\t{\tau}

\def\z{\zeta}
\def\D{\Delta}

\def\G{\Gamma}

\def\L{\Lambda}

\def\ch{{\cal H}}

\def\bo{{\raise-.5ex\hbox{\large$\Box$}}}               
\def\pa{\partial}                                       
\def\TH{{\raise.2ex\hbox{$\displaystyle \bigodot$}\mskip-4.7mu \llap H \;}}
\def\face{{\raise.2ex\hbox{$\displaystyle \bigodot$}\mskip-2.2mu \llap {$\ddot
        \smile$}}}                                      

\def\Tilde#1{\widetilde{#1}}                    
\def\Bar#1{\overline{#1}}                       
\def\abs#1{\left| #1\right|}                    
\def\leftrightarrowfill{$\mathsurround=0pt \mathord\leftarrow \mkern-6mu
        \cleaders\hbox{$\mkern-2mu \mathord- \mkern-2mu$}\hfill
        \mkern-6mu \mathord\rightarrow$}
\def\dvec#1{\vbox{\ialign{##\crcr
        \leftrightarrowfill\crcr\noalign{\kern-1pt\nointerlineskip}
        $\hfil\displaystyle{#1}\hfil$\crcr}}}           

\def\frac#1#2{{\textstyle{#1\over\vphantom2\smash{\raise.20ex
        \hbox{$\scriptstyle{#2}$}}}}}                   
\def\sfrac#1#2{{\vphantom1\smash{\lower.5ex\hbox{\small$#1$}}\over
        \vphantom1\smash{\raise.4ex\hbox{\small$#2$}}}} 
\def\bfrac#1#2{{\vphantom1\smash{\lower.5ex\hbox{$#1$}}\over
        \vphantom1\smash{\raise.3ex\hbox{$#2$}}}}       
\def\afrac#1#2{{\vphantom1\smash{\lower.5ex\hbox{$#1$}}\over#2}}    

\def\[{\lfloor{\hskip 0.35pt}\!\!\!\lceil}
\def\]{\rfloor{\hskip 0.35pt}\!\!\!\rceil}
\def\Lag{{\cal L}}
\def\du#1#2{_{#1}{}^{#2}}

\def\fracm#1#2{\hbox{\large{${\frac{{#1}}{{#2}}}$}}}

\def\un{\underline}
\def\fracmm#1#2{{{#1}\over{#2}}}

\def\low#1{{\raise -3pt\hbox{${\hskip 0.75pt}\!_{#1}$}}}

\def\Tilde#1{{\widetilde{#1}}\hskip 0.015in}

\newskip\humongous \humongous=0pt plus 1000pt minus 1000pt
\def\caja{\mathsurround=0pt}
\def\eqalign#1{\,\vcenter{\openup2\jot \caja
        \ialign{\strut \hfil$\displaystyle{##}$&$
        \displaystyle{{}##}$\hfil\crcr#1\crcr}}\,}
\newif\ifdtup

\def\pl#1#2#3{Phys.~Lett.~{\bf {#1}B} (19{#2}) #3}
\def\np#1#2#3{Nucl.~Phys.~{\bf B{#1}} (19{#2}) #3}

\def\cqg#1#2#3{Class.~and Quantum Grav.~{\bf {#1}} (19{#2}) #3}

\def\ibid#1#2#3{{\it ibid.}~{\bf {#1}} (19{#2}) #3}

\parskip=\medskipamount                 
\thicklines                         

\begin{document}
\thispagestyle{empty}

\noindent
\vskip1.3cm
\begin{center}

{\Large\bf Nonperturbative low-energy effective action \newline
                of the universal hypermultiplet}
\vglue.2in
                  Sergei V. Ketov

{\it          Department of Physics\\
           Tokyo Metropolitan University\\
          Hachioji, Tokyo 192--0397, Japan}\\
{\sl ketov@phys.metro-u.ac.jp}
\end{center}
\vglue.2in
\begin{center}
{\large\bf Abstract}
\end{center}

\begin{quote}
The non-perturbative (D-instanton corrected) low-energy effective action\\
of the universal hypermultiplet in the type-IIA string theory compactified \\
on a Calabi-Yau threefold is calculated in 4-dim, N=2 supergravity. The \\
action is given by the quaternionic non-linear sigma-model whose metric \\
is governed by the Eisenstein series $E_{3/2}$. The $U(1)\times U(1)$ isometry
and \\ 
the $SL(2,{\bf Z})$ modular invariance play the key role in our construction.
\end{quote}

A calculation of instanton corrections in compactified superstrings is crucial
for solving the fundamental problems of vacuum degeneracy and supersymmetry 
breaking. Some instanton corrections in the type-IIA superstring theory 
compactified on a {\it Calabi-Yau} (CY) threefold arise due to the Euclidean 
D2-branes wrapped about the CY special three-cycles \cite{bbs}. Being the BPS 
solutions to the Euclidean ten-dimensional supergravity equations of motion, 
these wrapped branes are localized in four uncompactified spacetime dimensions.
 They are called D-instantons. The D-instanton action is given by the volume 
of the supersymmetric 3-cycle on which a D2-brane is wrapped.  The 
supersymmetric cycles (by definition) minimize volume in their homology class. 

At the level of the {\it Low-Energy Effective Action} (LEEA), the effective 
field theory is given by the four-dimensional (4d), N=2 supergravity with 
some N=2 vector- and hyper-multiplets, whose structure is dictated by the CY 
cohomology, and whose moduli spaces are independent. The hypermultiplet sector
of the LEEA is described by a 4d, N=2 {\it Non-Linear Sigma-Model} (NLSM) 
with a quaternionic metric in the NLSM target (moduli)  space \cite{bw}. Any 
CY compactification gives rise to the so-called {\it Universal Hypermultiplet}
 (UH)  in 4d, which contains a dilaton amongst its field components. 

The bosonic part of the classical NLSM Lagrangian of UH in terms of the 
scalar fields  (a dilaton $\f$, an axion $D$, and a complex RR-scalar $C$) 
reads 
$$ - \Lag_{\rm FS} = (\pa_{\m}\f)^2 + e^{2\f}\abs{\pa_{\m}C}^2 
+ e^{4\f}(\pa_{\m}D +\fracm{i}{2}\bar{C}
\dvec{\pa_{\m}}C)^2~.~\eqno(1)$$
The metric of this NLSM is diffeomorphism-equivalent to the standard 
(quaternionic) Bergmann metric on $SU(2,1)/U(2)$. Equation (1) was derived 
\cite{fsh} by compactifying the 10d type-IIA supergravity on a CY threefold in
 the universal (UH) sector down to four spacetime timensions, $\m=0,1,2,3$. 
The metric (1) can be trusted as long as the string coupling is not strong. 

The absence of {\it perturbative} superstring (loop) corrections to the local 
UH metric was proved in ref.~\cite{one}. It is of considerable interest 
to calculate the UH {\it non}-perturbative NLSM metric by including all 
D-instanton corrections. Witten \cite{w1} showed that the D-instanton quantum 
corrections are given by powers of $e^{-1/g_{\rm string}}$,  where 
$g_{\rm string}$ is the type-II superstring coupling constant. An explicit 
non-perturbative solution to the UH metric, with all the D-instanton 
contributions included, was first constructed in ref.~\cite{sum}. We briefly 
describe this solution below.

Quantum corrections generically break all the continuous 
$SU(2,1)$ symmetries of the UH classical NLSM down to a discrete subgroup 
because of  charge quantization \cite{bbs}. As was demonstrated by Strominger 
\cite{one}, there is no non-trivial quaternionic deformation of the classical 
metric (1) in the superstring perturbation theory when {\it all} the 
Peccei-Quinn-type symmetries (with three real parameters $(\a,\b,\g)$), 
$$ D\to D+\a~,\quad C\to C+ \g -i\b~,\quad S\to S +2(\g+i\b)C+\g^2+\b^2~,
\eqno(2)$$
remain unbroken. However, {\it some} of the Peccei-Quinn-type symmetries 
(for example, the one associated with constant shifts of the RR scalar $C$) 
can be broken non-perturbatively (namely, by D-instantons) \cite{nbi}. The 
abelian isometry, associated with constant shifts of the axionic $D$-field, is
preserved when one considers D-instantons only, while five-brane instantons are
suppressed. The abelian rotations of the $C$ field are also preserved 
by D-instantons \cite{plb}. Our assumption of the $U(1)\times U(1)$ isometry 
of the non-perturbative UH moduli space metric is consistent with the known 
$U(1)\times U(1)$ symmetry of the Ooguri-Vafa solution \cite{ov} describing 
the D-instanton corrections to a matter hypermultiplet. Further support comes
from the type-IIB side, where the N=2 {\it double-tensor} multiplet version of
UH has to be used in the Euclidean path-integral approach to D-instantons 
\cite{tvv}. The Euclidean N=2 double-tensor multiplet action is bounded
from below, while its $U(1)\times U(1)$ symmetry (after dualization of two 
tensors into scalars) is protected against all quantum corrections \cite{tvv}.

As regards real {\it four}-dimensional
 quaternionic manifolds (relevant to the UH), they all have 
{\it Einstein-Weyl\/} geometry of {\it negative\/} scalar curvature, 
$$ W^-_{abcd}=0~,\qquad R_{ab}=-\fracmm{\L}{2}g_{ab}~,\eqno(3)$$  
where $W_{abcd}$ is the Weyl tensor, $R_{ab}$ is the Ricci tensor of the 
metric $g_{ab}$, and the constant $\L>0$ is proportional to the gravitational 
coupling constant. 

It is the theorem \cite{cp} that {\it any} four-dimensional Einstein-Weyl
metric (of non-vanishing scalar curvature) with two linearly independent 
Killing vectors can be written down in the form 
$$\eqalign{ 
ds^2_{\rm CP} ~=~ &  \fracmm{4\r^2(F^2_{\r}+F^2_{\h})-F^2}{4F^2}\,
\left(\fracmm{d\r^2+d\h^2}{\r^2}\right) \cr 
 & + \fracmm{ [(F-2\r F_{\r})\hat{\a}-2\r F_{\h}\hat{\b} ]^2 +[-2\r F_{\h}
\hat{\a}
+(F+2\r F_{\r})\hat{\b}]^2 }{F^2[4\r^2(F^2_{\r}+F^2_{\h})-F^2] }~,\cr}
\eqno(4)$$
in some local coordinates $(\r,\h,\q,\j)$ inside an open region of the 
half-space $\r>0$, where $\pa_{\q}$ and $\pa_{\j}$ are the two Killing 
vectors, the one-forms $\hat{\a}$ and $\hat{\b}$ are given by
$$ \hat{\a}= \sqrt{\r}\,d\q\quad {\rm and}\quad \hat{\b}=\fracmm{d\j 
+\h d\q}{\sqrt{\r}}~~,\eqno(5)$$
while the whole metric (4) is governed by the function 
(= {\it pre-potential}) $F(\r,\h)$ obeying a linear differential equation,
$$\D_{\ch}F \equiv \r^2\left(\pa^2_{\r}+\pa^2_{\h}\right)F =
\fracmm{3}{4}F~~.\eqno(6)$$

Equation (4) is thus a consequence of 4d, local N=2 supersymmetry. It is also 
remarkable and highly non-trivial that the {\it linear} equation (6) governs 
the solutions to the highly non-linear Einstein-Weyl system (3), given  
an $U(1)\times U(1)$ symmetry.

The linear equation (6) means that the pre-potential $F$ is a local 
eigenfunction (of the eigenvalue $3/4$) of the two-dimensional 
Laplace-Beltrami operator 
$$\D_{\ch} =\r^2(\pa^2_{\r}+\pa^2_{\h}) \eqno(7)$$
on the hyperbolic plane $\ch$ equipped with the metric 
$$ ds^2_{\ch}= \fracmm{1}{\r^2}( d\r^2 +d\h^2)~,\quad \r > 0~. 
\eqno(8)$$  
The metric (8) in the hyperbolic plane $\ch$ is invariant under the action 
of the isometry group $SL(2,{\bf R})$ isomorphic to $SU(1,1)$. The linearity 
of eq.~(6) allows a superposition of any two solutions to form yet another 
solution. In physical terms, this means that D-instantons have the cluster 
decomposition. 

For example, the pre-potential $F$ of the classical FS metric (1) is given by
 \cite{sum}
$$ F_{\rm FS}(\r,\h) = \fracmm{1}{\sqrt{\r}} \left\{
\sqrt{\h^2+\r^2}+ \fracmm{1}{4}\sqrt{ (\h-1)^2+\r^2} +
 \fracmm{1}{4}\sqrt{(\h+1)^2+\r^2}\right\}~~.\eqno(9)$$

The simplest (`basic') solutions to eq.~(6) are given by power functions,
$$ P_s(\r,\h) =\r^s~.\eqno(10)$$ 
Substituting eq.~(10) into eq.~(6) yields a quadratic equation,
$s(s-1)=3/4$, whose only solutions are 
$$ s_1=3/2\quad {\rm and}\quad s_2=-1/2~.\eqno(11)$$
Yet another `basic' solution to eq.~(6) is given by \cite{cp}
$$ \sqrt{ \r +\fracmm{\h^2}{\r}}~~.\eqno(12)$$

The $SL(2,{\bf Z})$ duality is supposed to be the exact symmetry of the 
compactified four-dimensional N=2 superstrings. The action of $SL(2,{\bf Z})$ 
in the hyperbolic plane $\ch$ is given by
$$ \t ~\to~ \hat{\g}\t=\fracmm{a\t +b}{c\t +d}~,\quad (a,b,c,d) 
\in {\bf Z}~,\quad ad-bc=1~,\eqno(13)$$
where the complex coordinate 
$$\t = \t_1 + i\t_2\equiv \h + i\r\eqno(14)$$ 
has been introduced. In terms of $\t$, the basic solutions take the form
$$ \fracmm{1}{\sqrt{\r}}=\fracmm{1}{\sqrt{{\rm Im}\,\t}} =
 \fracmm{1}{\sqrt{\t_2}}~,\eqno(15a)$$
$$\r^{3/2}=\left({\rm Im}\,\t\right)^{3/2}=(\t_2)^{3/2}~,\eqno(15b)$$
and
$$\sqrt{\r +\fracmm{\h^2}{\r}}=\sqrt{\fracmm{\t\bar{\t}}{{\rm Im}\,\t}} =
\fracmm{\abs{\t}}{\sqrt{\t_2}}~,\eqno(15c)$$
where $\abs{\t}^2=\t\bar{\t}=\t_1^2+\t^2_2=\r^2+\h^2$.

The most general $SL(2,{\bf Z})$-invariant solution to eq.~(6) is obtained 
by applying the discrete duality transformations (13) to all basic solutions 
(15) and summing over all these transformations (modulo a stability 
subgroup of each basic solution). 

As regards the power solutions (15a) and (15b), summing them over the
$SL(2,{\bf Z})$ gives rise to the non-holomorphic modular (automorphic) forms 
called the Eisenstein series \cite{ter}, 
$$E_s(\t,\bar{\t})=\sum_{\hat{\g}\in \Tilde{SL(2,{\bf Z})}}
P_s(\hat{\g}\t,\Bar{\hat{\g}\t})~,\eqno(16)$$
with $s=-1/2$ and $s=3/2$, respectively. To be well defined, the sum in 
eq.~(16) has to be limited to the quotient  
$\Tilde{SL(2,{\bf Z})}=SL(2,{\bf Z}/\G_{\infty}$ with the stabilizer 
$$  \G_{\infty} =\left\{ \left(\begin{array}{cc} \pm 1 & * \\
0 & \pm 1 \end{array} \right)\in SL(2,{\bf Z})\right\}~.\eqno(17)$$

In the case of ${\rm Re}\,s>1$, the Eisenstein series (16) is given by 
\cite{ter}
$$ E_{s}(\t,\bar{\t})
=\fracmm{1}{2}\t_2^s\sum_{(p,n)=1}\fracmm{1}{\abs{p+n\t}^{2s}}=
\t_2^s +\t_2^s\sum_{(p,n)=1\atop n\geq 1}\fracmm{1}{\abs{p+n\t}^{2s}}~~,
\eqno(18)$$
where $(p,n)$ is the greatest common divisor of $p$ and $n$. The infinite sum
 (18) can be interpreted as the contributions from the D-instantons of
discrete energy $p$ and discrete charge $n$ (see below). The duality-invariant 
sum of the basic solutions (15b) is thus given by the Eisenstein series 
(16) of $s=3/2$.

The Eisenstein series (as the functions of $s$) satisfy the functional equation
\cite{ter}
$$ \L(s)E_s(\t,\bar{\t})=\L(1-s)E_{1-s}(\t,\bar{\t})~,\eqno(19)$$
where
$$\L(s) =\p^{-s}\G(s)\z(2s)~.\eqno(20)$$ 
In particular, in the case of $s=3/2$, eqs.~(19) and (20) imply
$$ E_{-1/2}(\t,\bar{\t}) = \fracmm{3\z(3)}{\p^2} E_{3/2}(\t,\bar{\t})~.
\eqno(21)$$
Therefore, the duality-invariant sum over the basic solutions (15a) yields
{\it the same} result as that of eq.~(15b), namely, the Eisenstein series 
$E_{3/2}(\t,\bar{\t})$.

As regards the sum of (15c) over the $SL(2,{\bf Z})$, it also gives rise to 
{\it the same} function proportional to $E_{3/2}(\t,\bar{\t})$. This can be 
most easily seen by noticing that the $SL(2,{\bf Z})$ transformations are 
generated by the T-duality transformation,  $\h\to \h+1$, and the S-duality 
transformation, ${\rm S}:~~\t ~\to~ -\fracmm{1}{\t}$. When being applied 
to the basic solution (15a), the S-duality yields the basic solution (15c):
$$ \fracmm{1}{\sqrt{\r}}=\fracmm{1}{\sqrt{{\rm Im}\,\t}} 
~\stackrel{\rm S}{\longrightarrow}~ \fracmm{\abs{\t}}{\sqrt{{\rm Im}\,\t}}= 
\sqrt{\r +\fracmm{\h^2}{\r}}~~~.\eqno(22)$$

The Eisenstein series $E_s(\r,\h)$ has a Fourier series expansion \cite{ter}
$$\eqalign{
\L(s)E_s(\r,\h) ~=~& \r^s\L(s) + \r^{1-s}\L(1-s) \cr
~& +2\r^{1/2}\sum_{m\neq 0}\abs{m}^{s-1/2}\s_{1-2s}(m)K_{s-1/2}(2\p\abs{m}\r)
e^{2\p im\h}~~.\cr}\eqno(23)$$
Here $\s_s(m)$ is the so-called divisor function
$$ \s_s(m)=\sum_{0<d|m}d^s~,\eqno(24)$$
where the sum runs over all positive divisors $d$ of $m$. In the case of
$s=3/2$, eq.~(23) can be put into the form 
$$ 4\p E_{3/2}(\r,\h)= 2\z(3) \r^{3/2} +\fracmm{2\p^2}{3}
\r^{-1/2} + 8\p\r^{1/2}\sum_{m\neq 0 \atop n\geq 1} \abs{\fracmm{m}{n}}
e^{2\p imn\h}K_1(2\p\abs{mn}\r)~,\eqno(25)$$
where $\z(3)= \sum_{m>0}(1/m)^3$. The asymptotical expansion of the function
(25) for large $\r$ is given by
$$\eqalign{
4\p E_{3/2}(\r,\h) = &  2\z(3)\r^{3/2} +\fracmm{2\p^2}{3}\r^{-1/2} 
+4\p^{3/2}\sum_{m,n\geq 1}\left(\fracmm{m}{n^3}\right)^{1/2}\left[
e^{2\p i mn(\h+i\r)}+e^{-2\p i mn(\h-i\r)}\right] \cr
  & \times \left[ 1 + \sum^{\infty}_{k=1}\fracmm{\G(k-1/2)}{\G(-k-1/2)}\,
\fracmm{1}{(4\p mn\r)^k}\right]~. \cr}\eqno(26)$$

We conclude that the infinite $SL(2,{\bf Z})$-invariant D-instanton sum of all
basic solutions (15) is always proportional to the Eisenstein series 
$E_{3/2}(\t,\bar{\t})$. Hence, the potential $F$ of the UH metric with all the
$D$-instanton contributions included is proportional to $E_{3/2}(\t,\bar{\t})$
too. Our quaternionic D-instanton sum in the form of the Eisenstein series is 
a consequence of local N=2 supersymmetry, toric isometry $U(1)\times U(1)$ 
and $SL(2,{\bf Z})$ duality.

The Eisenstein series (25) also appears in the exact non-perturbative 
description of the $R^4$ couplings in the ten-dimensional type-IIB 
superstrings \cite{gg,gg2}. When being expanded in the form (26), it
amounts to the infinite sum of the tree level, one-loop, and D-instanton 
contributions, respectively. It is therefore conceivable that the 
non-perturbative UH moduli space metric with the D-instanton contributions is 
completely determined by the exact $R^4$ couplings in the ten-dimensional 
type-IIB superstrings, similarly to the known case of the one-loop (locally 
trivial) contribution \cite{cand}. The exact $R^4$ couplings in the 
ten-dimensional type-IIB superstrings are dictated by {\it the same} equation 
(3.17) \cite{sethi,boris}. 
\vglue.2in

\end{document}